\documentclass[]{article}

\usepackage{amssymb,latexsym,amsmath}     
\usepackage{authblk}
\usepackage[utf8]{inputenc}
\usepackage{dsfont}
\usepackage[utf8]{inputenc}
\usepackage{graphicx}
\usepackage{subfig}
\usepackage{graphicx}
\usepackage{xcolor}

\usepackage[colorlinks = true,
linkcolor = blue,
urlcolor  = blue,
citecolor = blue,
anchorcolor = blue]{hyperref}
\usepackage[numbers,square,sort&compress]{natbib}

\usepackage{authblk}
\usepackage{cancel}

\usepackage{enumitem}

\begin{document}
	\title{\bf Analogue gravity and the island prescription} 
\author[ ]{Shahrokh Parvizi}
\author[ ]{Mojtaba Shahbazi}

\affil[ ]{ Department of Physics, School of Sciences,
	Tarbiat Modares University, P.O.Box 14155-4838, Tehran, Iran} 
	\affil[ ]{\textit {Email: \href{mailto:parvizi@modares.ac.ir}{parvizi@modares.ac.ir}, \href{mailto:mojtaba.shahbazi@modares.ac.ir}{mojtaba.shahbazi@modares.ac.ir}}}
	\date{\today}                     
	\setcounter{Maxaffil}{0}
	\renewcommand\Affilfont{\itshape\small}
	
	\maketitle
\begin{abstract}
Analogue gravity succeeded to simulate Hawking radiation and test it in laboratories. In this setting, the black hole is simulated by an area in a fluid, say water, where no sound wave can escape the event horizon and phonon oscillations are detected as Hawking radiation. This means that the analogue simulations can provide an alternative description, and consequently, a new insight to the high energy physics problems. Now it would be interesting to see what information loss means and how island prescription is interpreted in water experiment. In this paper we show that the analogue of information loss is the loss of momentum per unit mass of the fluid over the horizon and maintaining the momentum loss leads to the island prescription.
\end{abstract}
\section{Introduction}
The key question of the information paradox is if the formation and the evaporation of black holes follow a unitary evolution. Equivalently, whether the entanglement entropy of the Hawking radiation follows the so called ``Page curve", initially rises and then backs down \cite{page}. Hawking's original calculations show that the entanglement entropy of the radiation increases monotonically. However, if one goes with the AdS/CFT dual of the black hole, a quantum mechanical description, it is believed that the Page curve is realized because of the unitary evolution. That is a question how we can find the Page curve in the gravitational side.
Ryu-Takayanagi (RT) formula \cite{RT} suggests how to compute the entanglement entropy (EE) in the gravitational side by computing the area of an extremized surface in the bulk, this surface is homologous to the subregion on the boundary. It has been shown that RT formula lacks the quantum corrections \cite{malda} and one can instead find quantm extremized surface (QES) or use the generalized entropy to take into account the quantum corrections \cite{penington}.
The QES extremizes the generalized entropy which is defined as:
\begin{equation}
S\sim min\Big(\frac{Area}{4G_N}+S_{outside}\Big)\label{gen}
\end{equation}
where $S_{outside}$ is the entropy of the bulk fields outside the QES. The generalized entropy predicts a phase transition in the EE of the radiation; however, it does not reproduce the Page curve \cite{bulkfield}. By the fact that, there is a phase transition in the EE of the radiation via QES, it seems that inside of the black holes could contribute to find the Page curve for the EE of the radiation because this phase transition is due to the penetration of QES behind the horizon. Inspired by this fact, the island prescription has been introduced \cite{almehri}:
\begin{equation}
S\sim min\Big(\frac{Area(\partial \Sigma_{island})}{4G_N}+S_{outside~ matter}(\Sigma_{radiation}\cup \Sigma_{island})\Big) \label{island}
\end{equation}
where $\Sigma_{island}$ is a surface inside the black hole and $\Sigma_{radiation}$ the surface where the Hawking radiation lives as Fig. \ref{ip} depicts. The UV divergences of the $S_{matter}$ is absorbed by a renormalization of the Newtonian constant:
$$\frac{1}{4G_N}\rightarrow \frac{1}{4G_N}-\frac{1}{\epsilon^2}$$
To apply this approach for computing the entropy of the Hawking radiation, 
one should look for the holographic dual to the quantum field theory on a curved background. The holographic dual to this scenario is a one extra dimensional gravitational theory with an end-of-the-world brane which is a brane attached to a cut-off radius \cite{gubser}. Due to the fact that large black holes in an AdS space-time reach thermal equilibrium, to proceed the evaporation of black holes, a non-gravitational thermal bath is attached to the location of the brane with a transparent boundary condition to let the radiation pass through the boundary. Put it another way, there is a CFT in flat background as a bath coupled to a black hole. This procedure is called a doubly-holographic picture \cite{almehri}.

Although the island formula seems to be ad hoc, the formula has been confirmed by the path integral computation and the replica trick in lower dimensional gravities \cite{stanford, maldacena}. The Page curve has been retrieved by \eqref{island}, which means that the information paradox has been resolved by an AdS/CFT approach. Although, it is not clear how direct computations can deal with the information paradox, it is believed that a quantum gravity theory has a proper resolution.

To better understand the information paradox and see how the island prescription preserves the information, we are going to simulate the black hole in analogue gravity, particularly in a fluid and see what happens when there is information loss. It has been shown that a perturbation in
the analogue models of gravities share the same equations of motion as a quantum field theory in a curved background \cite{analogue}. This means that we can simulate a quantum field theory in a curved background by its analogue model such as a fluid and find black holes solutions in these models which are called ``dumb hole". Interestingly, as the same as the black hole radiation in the presence of a quantum field theory, dumb holes radiate as well \cite{unruh}. The radiation, here, is the oscillation of the medium or phonon in the fluid. There are some experimental evidences for the phonon oscillation in laboratories which means that there are experimental confirmations of the Hawking radiation in the analogue models such as fluid \cite{fluid} and atomic Bose–Einstein condensate \cite{be}. In this manner, it seems reasonable to find the analogue of the information loss and show that how the island prescription preserves the information.
In the proceeding sections, it will be shown that the analogue of the information loss is the momentum loss of the fluid over the dumb hole horizon, put it another way, if one translates the information loss in the analogue systems and simulates it in these systems it turns into the momentum loss over the horizon in the fluid; however, in real experiments of the analogue gravity there is a water pump or something that maintains the momentum and makes the steady flow in the fluid, we will see that the information paradox in the analogue gravity arises due to the negligence of the momentum into the horizon maintained by the water pump. If one takes into account the effect of pump maintenance by the Newton's second law, the island prescription is retrieved and here the island prescription is no longer ad hoc. In other words, analogue gravity by simulation of the gravitational phenomenon provides a method to solve problems.
It should be noted that the analogue gravity is considered in two ways:
\begin{itemize} 
\item Firstly, it could be considered as a simulation of classical fields in a curved background where the ratio of ingoing and outgoing modes of the the classical field is measurable as in \cite{fluid,complexity}. 
\item Secondly, quantization of phonon in a condensed matter system simulates the quantization of fields in the curved background such as \cite{be,ge1,ge2,libe1,libe}.
\end{itemize} 
In this paper we follow the first approach in holography. It means that for computing the EE of the Hawking radiation we go into the the bulk theory of the Hawking radiation which is a classical gravity and finding an area of a surface gives the EE of the Hawking radiation. This bulk theory could be simulated by the analogue gravity, what we do in the following. It is worth mentioning that the bulk theory (in one extra dimensional theory) of the Hawking radiation is an eternal two-sided static black hole coupled to a non-gravitational bath which means that despite the evaporating black hole on the boundary, the horizon of the dual evaporating black hole in the bulk does not shrink.

The rest of the paper is as follows: in the next section, we review the calculations of entanglement entropy (EE) of the radiation with and with out the island prescription. In section 3, we find the analogue of the EE and the information loss in terms of fluid parameters and explain what it is missed when the information loss arises and how the island prescription takes that into account.
\section{Entanglement entropy of the Hawking radiation}
\begin{figure}[!]
\centering
\includegraphics[width=10cm]{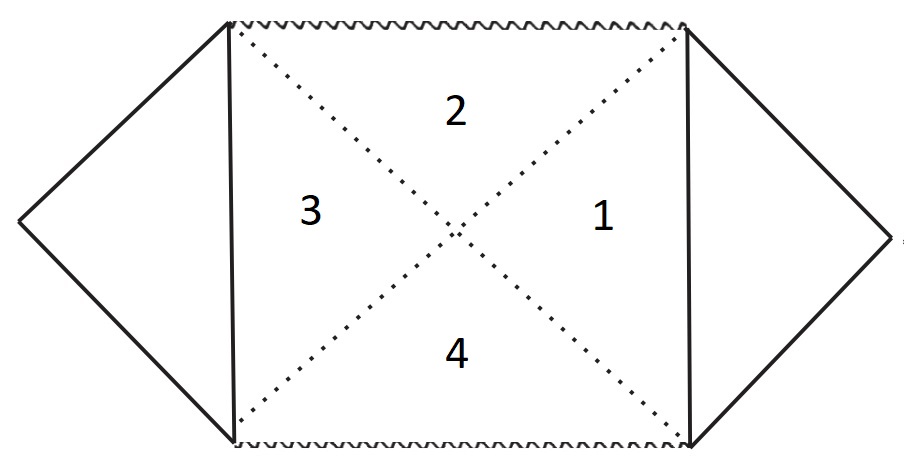}
\caption{The Penrose diagram of a two-sided black hole coupled to a non-gravitational bath.}\label{ads}
\end{figure}

To compute the EE of the radiation we should follow the island formula \eqref{island} keep on with the matter contribution for the times before the Page time and with island for the times later than the Page time. The computations are conducted for an asymptotic AdS Schwarzschild black hole in $d+1$ dimensions. In asymptotic flat spacetimes with massless gravitons it has been found that due to the long-range gravity and non-locality the algebra of operators associated with the island entanglement wedge\footnote{The bulk causal diamond constructed in the interior of the extremized surface is called the entanglement wedge of the boundary region} leads to an inconsistency even in the asymptotic flat spacetimes coupled to a non-gravitational bath, then it is not allowed to use the island formula in these spacetimes. Put it differently, the demarcation of the island QES presupposes the locality which is absent in these cases \cite{randall}.
The line element of an asymptotic AdS Schwarzschild black hole in $d+1$ dimensions is written as:
\begin{equation}
ds^2=\omega^2\Big(-fdt^2+\frac{dr^2}{f}+r^2d\sigma_{d-1}^2\Big) \label{sch}
\end{equation}
where $\omega$ is a conformal factor for later convenience. The computation of EE of the Hawking radiation is easily done if the line element written in the Kruskal coordinates. The Kruskal coordinates for different wedges of the space-time (see Fig. \ref{ads}) are introduced as:
\begin{align}
1:~&u=\frac{e^{k(r^*+t_{\scriptscriptstyle R})}}{k}~~~v=\frac{-e^{k(r^*-t_{\scriptscriptstyle R})}}{k}\\
2:~&u=\frac{e^{k(r^*+t_{\scriptscriptstyle R})}}{k}~~~v=\frac{e^{k(r^*-t_{\scriptscriptstyle R})}}{k}\\
3:~&u=\frac{-e^{k(r^*-t_{\scriptscriptstyle R})}}{k}~~~v=\frac{e^{k(r^*+t_{\scriptscriptstyle R})}}{k}\\
4:~&u=\frac{-e^{k(r^*+t_{\scriptscriptstyle R})}}{k}~~~v=\frac{-e^{k(r^*-t_{\scriptscriptstyle R})}}{k}
\end{align}
where $r^*(r):=\int^r\frac{dr}{f}$, and $k=\frac{f'(R)}{2}$ is the surface gravity at the horizon radius $R$.
The line element in Kruskal coordinates is written as:
\begin{equation}
ds^2=\omega^2\Big(\frac{f}{k^2uv}dudv+r^2d\sigma_{d-1}^2\Big):=\Omega dudv+r^2\omega^2d\sigma_{d-1}^2 \label{kru}
\end{equation}
in which $\Omega=-\omega^2fe^{-2kr^*}$. In the next section, we are going to compute the EE of the Hawking radiation for the early times by the island formula \eqref{island}.
\subsection{Without the island}
It has been shown that the EE of the Hawking radiation before the Page time could be computed by \eqref{island} without the island. Using the s-wave approximation, the EE of the matter in \eqref{island}, in background of a line element in the form \eqref{kru}, is given by \cite{dist}:
\begin{equation}
S_{rad.}=\frac{\mathcal{C}}{6}logD^2(b_1,b_3)=\frac{\mathcal{C}}{6}log\Big|\bigg(u(b_1)-u(b_3)\bigg)\bigg(v(b_1)-v(b_3)\bigg)\sqrt{\Omega(b_1)\Omega(b_3)}\Big|
\end{equation}
\begin{figure}[h]
\centering
\includegraphics[width=10cm]{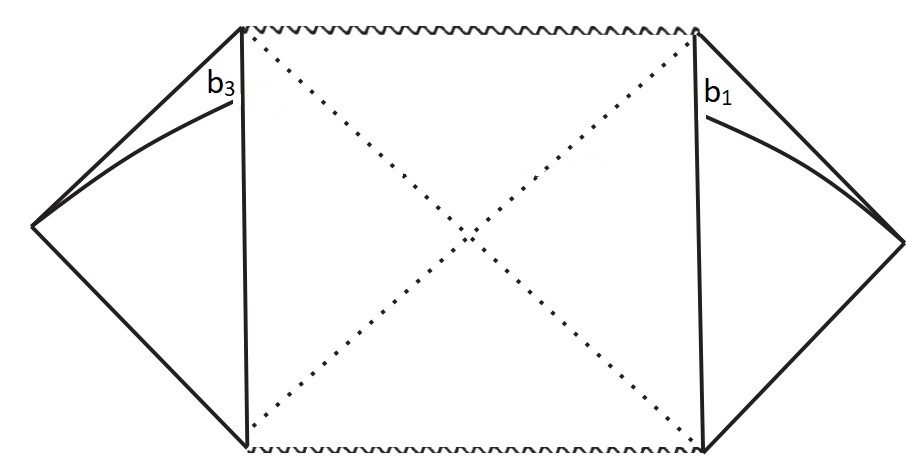}
\caption{The solid lines terminated to $b$'s coordinates are the subsystem of the CFT where the Hawking radiation is collected.} \label{adswoi}
\end{figure}
where $\mathcal{C}$ is the central charge of the CFT, $b_1$ and $b_3$ are the radial coordinates of the subsystem collecting the Hawking radiation as in Fig \ref{adswoi}. We take
$r^*(b_1):=r^*_1$, $r^*(b_3):=r^*_3$ and because the two subsystems are located in the same Cauchy surface $t_1=t_3=t$ and set $r^*_1=r^*_3=r^*_b$.
\begin{align}
D^2(b_1,b_3)&=\Big|\bigg(-\frac{e^{k(r^*_1-t)}}{k}-\frac{e^{k(r^*_3+t)}}{k}\bigg)\bigg(\frac{e^{k(r^*_1+t)}}{k}+\frac{e^{k(r^*_3-t)}}{k}\bigg)\sqrt{\omega^4(b)f^2(b)e^{-4kr^*_b}}\Big|\\ \label{distance}
&=\frac{4\omega^2(b)f(b)}{k^2}\cosh^2(kt)
\end{align}
\begin{equation}
S_{rad.}=\frac{\mathcal{C}}{6}log\Big(\frac{4\omega^2(b)f(b)}{k^2}\cosh^2(kt)\Big) \label{woi}
\end{equation}
at late times:
\begin{align}
S_{rad.}&\approx \frac{\mathcal{C}}{3}kt +\frac{\mathcal{C}}{6}log\Big(\frac{4\omega^2(b)f(b)}{k^2}\Big)\\
&= \frac{\mathcal{C}t}{6}f'(R) +\frac{\mathcal{C}}{6}log\Big(\frac{4\omega^2(b)f(b)}{k^2}\Big)\label{eewoi}
\end{align}
If the EE of the Hawking radiation is computed by \eqref{eewoi}, it is evident that the EE increases monotonically and the Page curve is not followed. Indeed, Hawking's original computations follow \eqref{eewoi} even after the Page time.
Notice that the EE of the Hawking radiation at $t=0$ is:
\begin{equation}
S_{rad.}(t=0):=S_{initial}=\frac{\mathcal{C}}{6}log\Big(\frac{4\omega^2(b)f(b)}{k^2}\Big)
\end{equation}
It means that if the Page curve was supposed to be followed then the EE of the Hawking radiation at late times should be equal to $S_{rad.}(t=0)$. Then the first term in the right-hand side of \eqref{eewoi}, $\frac{\mathcal{C}t}{6}f'(R)$ is responsible for the information loss.
In the next section, the fluid interpretation of the EE of the radiation in analogue gravity is going to be considered.
\section{Analogue gravity}
Analogue models for gravity simulate the gravitational phenomenon in condensed matter systems. To find the analogue model for the gravity we should stick to the following instruction (for a quick review see the appendix \ref{analogueapp}).
If we use the transformation $t=t'\alpha+\beta$ for $\beta=\int \frac{\sqrt{1-f}}{f}$, then the line element \eqref{sch} could be written as:
\begin{equation}
ds^2=\omega^2\Big(-\alpha^2fdt'^2+2\alpha\sqrt{1-f}dt'dr+dr^2+r^2d\Omega_{d-1}^2\Big)
\end{equation}
By comparing with the acoustic line element (see appendix \ref{analogueapp}):
\begin{equation}
ds^2=\Big(\frac{\rho}{\kappa c}\Big)^{\frac{2}{n-1}}\Big(-c^2dt^2+\delta_{ij}(dx^i-v^idt)(dx^j-v^jdt)\Big)
\end{equation}
the acoustic parameters are read:
\begin{align}
&t-r~component:~~v=\alpha\sqrt{1-f}\label{vel}\\
&r-r~component:~~\rho=\kappa c\omega^{n-1} \label{den}
\end{align}
where $\alpha=c$. To satisfy the continuity equation $\nabla.(\rho v)=0$ in spherical coordinates:
\begin{align}
&\nabla.(\rho v)=0\rightarrow \partial_r(r^{n-1}\rho v)=\partial_r(r^{n-1}\kappa \omega^{n-1}\sqrt{1-f})=0
\end{align}
then
\begin{align}
&r^{n-1}\kappa \omega^{n-1}\sqrt{1-f}=cte.:=\gamma
\end{align}
and it follows that
\begin{align}
\gamma=c R^{n-1}\rho_{\scriptscriptstyle R}\;,\qquad \omega=\frac{1}{r}\Big(\frac{\gamma}{\kappa\sqrt{1-f}}\Big)^{\frac{1}{n-1}}\;. \label{omega}
\end{align}
where $\rho_{\scriptscriptstyle R}:=\rho (R)$.
To satisfy Euler equation, $\rho \big(\partial_tv+(v.\nabla)v\big)=\mathcal{F}$ where $\mathcal{F}$ is the force per volume or force density, we find in spherical coordinate:
\begin{align}
\rho v_r\partial_rv_r=\mathcal{F}_r\;.
\end{align}
By putting the parameters \eqref{vel} and \eqref{den} in Euler eq.:
\begin{align}
\mathcal{F}_r=-\frac{1}{2}c^2\rho f'=-\frac{\gamma c^3f'}{2r^{n-1}\sqrt{1-f}}
\end{align}
then the force density over the horizon is given by:
\begin{align}
\mathcal{F}_{\scriptscriptstyle R}:=\mathcal{F}_r(R)=-\frac{\gamma c^3f'(R)}{2R^{n-1}}\;.
\end{align}
The force density could be written in terms of the mass density \eqref{omega} over the horizon:
\begin{align}
\mathcal{F}_{\scriptscriptstyle R}=-\frac{1}{2}c^2\rho_{\scriptscriptstyle R} f'(R)\;.
\end{align}
We therefore may consider the quantity
\begin{align}\label{lossho}
-\frac{1}{2}c^2 f'(R)&=\frac{\mathcal{F}_{\scriptscriptstyle R}}{\rho_{\scriptscriptstyle R}}=\frac{F}{M}=\frac{\frac{dP}{dt}}{M}  
\end{align}
as the momentum transfer per unit mass over the horizon where $F$ is the force inserted on the control volume in the fluid and $P$ is the momentum of the fluid\footnote{$M$, the mass of the control volume is the mass of the fluid contained behind the dumb hole horizon.}. In other words, $\mathcal{F}_{\scriptscriptstyle R}/\rho_{\scriptscriptstyle R}$ represents the momentum per unit mass over the horizon.
Recalling the EE of the Hawking radiation \eqref{eewoi} and the term responsible for the information loss $\frac{\mathcal{C}t}{6}f'(R)$ which by \eqref{lossho} can be written as 
\begin{equation} 
\frac{\mathcal{C}t}{6}f'(R)=-\frac{\mathcal{C}t}{3c^2}\frac{\mathcal{F}_{\scriptscriptstyle R}}{\rho_{\scriptscriptstyle R}}=-\frac{\mathcal{C}}{3c^2}\frac{t\frac{dP}{dt}}{M}=-\frac{\mathcal{C}}{3c^2}\frac{\Delta P}{M},
\end{equation}
which means that the analogue of the information loss is represented by the momentum loss per unit mass over the horizon. Put it another way, the information loss (IL) arises due to the momentum loss over the horizon. Now \eqref{eewoi} could be rewritten as:
\begin{equation}
S_{rad.}= -\frac{\mathcal{C}}{3c^2}\frac{\Delta P}{M}\Big|_{out}+\frac{\mathcal{C}}{6}log\Big(\frac{4\omega^2(b)f(b)}{k^2}\Big)\label{mtran}
\end{equation}
To be concrete, the momentum flow per unit mass over the horizon, $\frac{\Delta P}{M}\Big|_{out}$ is equivalent to $\frac{3 c^2}{\mathcal{C}}S_{loss}$, where $S_{loss}$ is the portion of EE to be lost. Roughly speaking, $\frac{3 c^2}{\mathcal{C}}S_{rad.}$ could be interpreted as the momentum transfer per unit mass.
\subsection{Resolution of IL in AG}
In the previous section, it has been seen that the analogue of the IL is equivalent to the momentum transfer over the horizon in analogue gravity. In this section, the resolution of the IL is going to be considered.
It is worth mentioning that the shrinkage of the horizon radius during Hawking radiation is a quantum effect not a classical one \cite{libe}. Due to the fact that in this paper we are concerned with the classical simulation then the horizon radius is fixed and as a consequence the velocity of the fluid simulating the black hole is a steady flow as in \cite{fluid} and it does not require to take into account the shrinkage effect as in \cite{libe}. Besides that, we got the EE of the radiation by holographic computations (the island formula) in which there is no shrinkage of the horizon radius. In other words, in this paper we find the analogue gravity model that simulates the black hole and the EE of the radiation in the bulk of holography\footnote{Island formula which is a proposal for computing the EE of the radiation through the bulk computes the EE of the subregion boundary \cite{almehri}. What we do in this paper is the simulation of the bulk theory which is a classical gravitational theory, in our computations a two-sided eternal black hole coupled to non-gravitational bath. In this setting the horizon radius is fixed.}.

\begin{figure}[!]
\centering
\includegraphics[width=7cm]{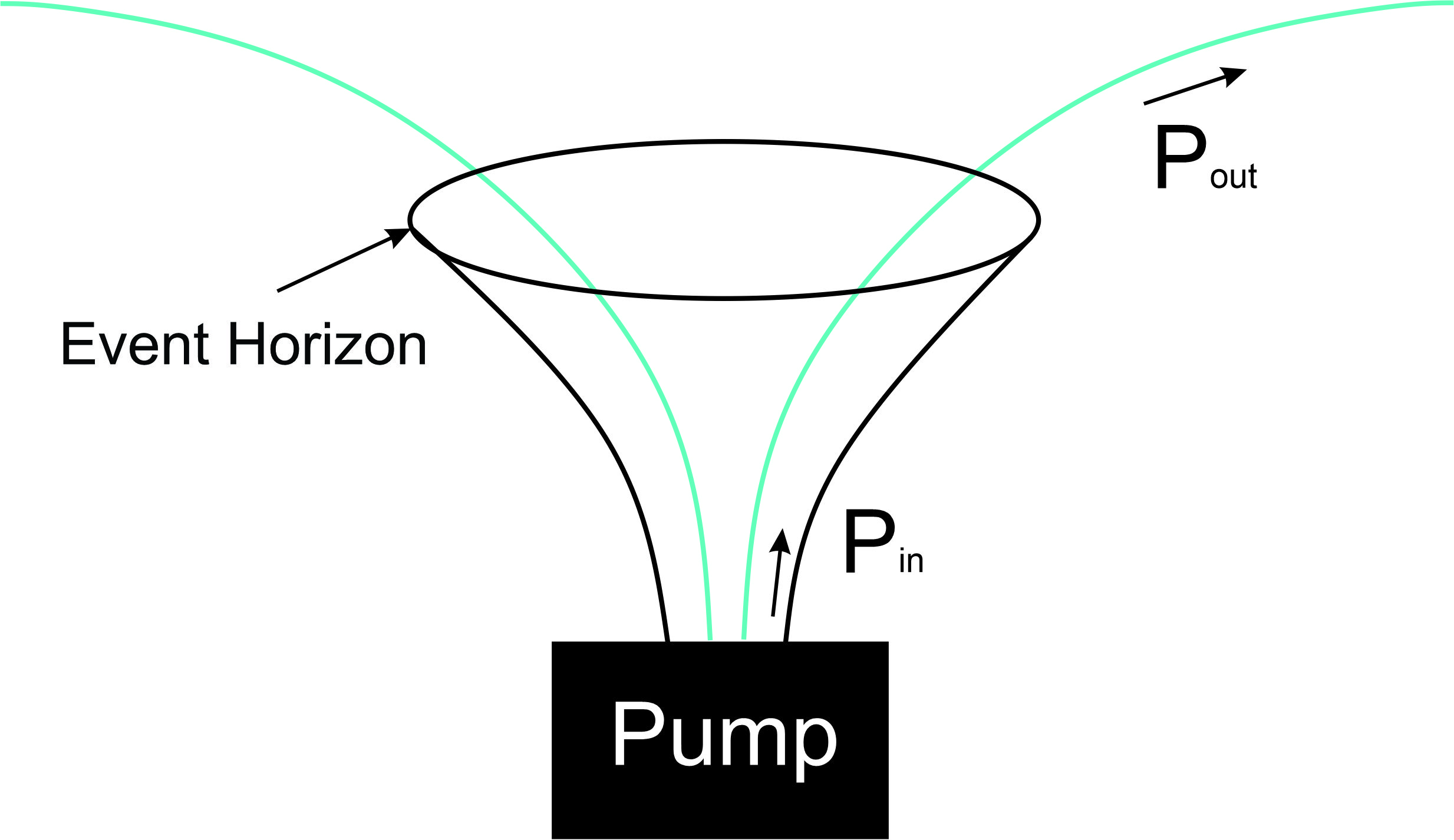}
\caption{The control volume is the interior of the dumb hole and the blue lines show the flow of the water.} \label{pump}
\end{figure}

In real experiments of analogue gravity to maintain the steady flow of the fluid momentum, one uses a water pump\footnote{An experimental set-up to produce a dumb hole has been shown in \cite{fluid} and the role of water pump is to maintain the steady flow in that setting. However, in a real-life setting such as a river, the gravitational potential plays the role of the water pump}. In Fig \ref{pump} which is a schematic diagram of an analogue gravity, a pump maintains the water behind the event horizon. If one chooses the control volume as the interior of the dumb hole, to have a fixed horizon it requires vanishing the net force inserted into the interior of the dumb hole or equivalently, the momentum rate of the interior of the dumb hole should be vanished:
\begin{align}
&\sum F_{net}=0 \rightarrow \frac{dP}{dt}\Big|_{in}-\frac{dP}{dt}\Big|_{out}=0\rightarrow \frac{t}{M}\frac{dP}{dt}\Big|_{in}-\frac{t}{M}\frac{dP}{dt}\Big|_{out}=0
\end{align}
In the previous section, it has been found out that in the occasion of IL there is a momentum flow outwards from the horizon, then to make the net force vanishing, there should be a momentum flow inwards to the interior of the dumb hole with the same value of outwards momentum. Put it differently, there is the momentum loss because of the fact the inwards momentum is neglected. If one adds the inwards momentum to the computations of the previous section then the EE of the Hawking radiation is rewritten as:
\begin{align}
S_{rad.}&= \frac{\mathcal{C}}{3 c^2}\frac{\Delta P}{M}\Big|_{in}-\frac{\mathcal{C}}{3 c^2}\frac{\Delta P}{M}\Big|_{out} +\frac{\mathcal{C}}{6}log\Big(\frac{4\omega^2(b)f(b)}{k^2}\Big) \label{extra} \\
&=\frac{\mathcal{C}}{6}log\Big(\frac{4\omega^2(b)f(b)}{k^2}\Big)\label{anaee}
\end{align}
The point is that in the previous section it has been revealed that $\frac{3 c^2}{\mathcal{C}}S_{loss}=-\frac{\Delta P_{out}}{M}$, so it is reasonable to think that $\frac{3 c^2}{\mathcal{C}}S_{compensation}=+\frac{\Delta P_{in}}{M}$ where $S_{rad.}=S_{loss}+S_{compensation}+S_{initial}$.
The remaining term in \eqref{anaee} could be shown that is related to the island prescription. For large $b$, we have:
\begin{align}
log\Big(\frac{4\omega^2(b)f(b)}{k^2}\Big) \approx log\Big(\omega^2(b)f(b)\Big)\approx logD^2(a_r,b_r)
\end{align}
which means that:
\begin{align}
S_{rad.}\approx \frac{\mathcal{C}}{6} log\Big(\omega^2(b)f(b)\Big) \label{app}
\end{align}
\begin{figure}[h]
\centering
\includegraphics[width=10cm]{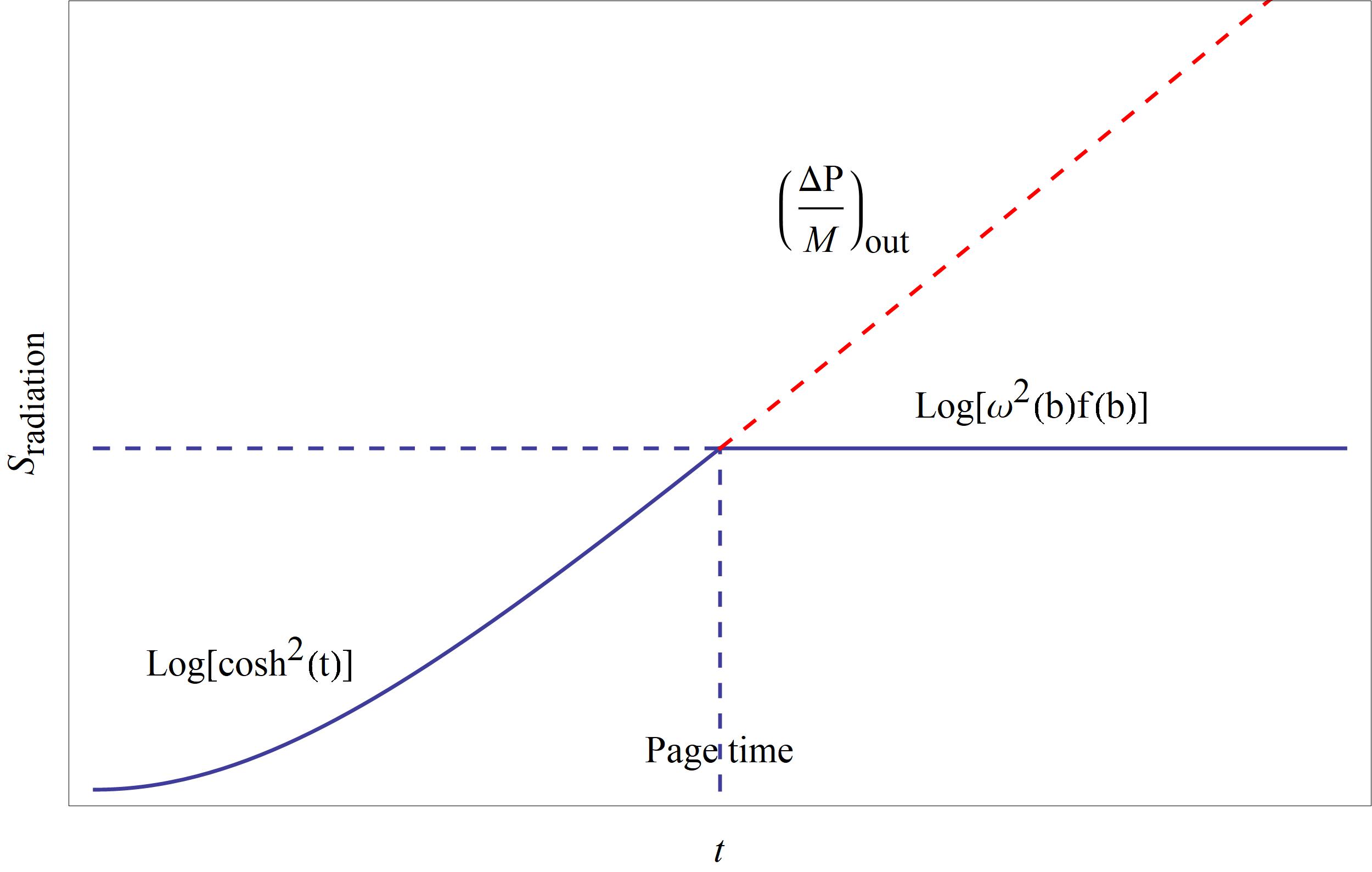}
\caption{EE of the Hawking radiation in terms of time. The solid line is the EE of the Hawking radiation before and after the Page time and the red dashed line is the Hawking's calculation where the information paradox arises. The constants in \eqref{woi} and \eqref{app} are chosen to be 1.} \label{pc}
\end{figure}
The EE of the Hawking radiation is drawn in Fig. \ref{pc}. In the next section, we show that how \eqref{app} is equal to the island prescription.

\subsection{EE with the island}
\begin{figure}[h]
\centering
\includegraphics[width=10cm]{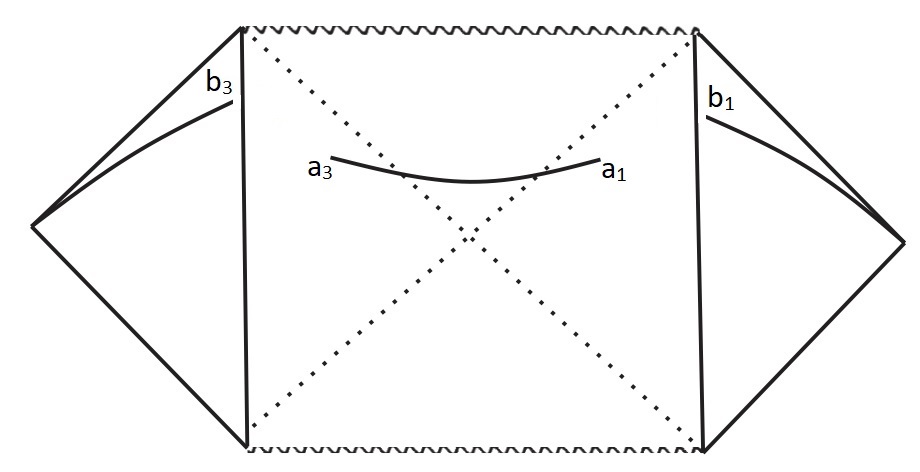}
\caption{$a$'s coordinates are the radial coordinates of the island.} \label{ip}
\end{figure}

The EE of the Hawking radiation by the island prescription, \eqref{island} is given by:
\begin{equation}
S_{rad.}=\frac{A}{2G}+\frac{2\mathcal{C}}{6}logD^2(a_1,b_1)=\frac{A}{2G}+\frac{2\mathcal{C}}{6}log\Big|\bigg(u(a_1)-u(b_1)\bigg)\bigg(v(a_1)-v(b_1)\bigg)\sqrt{\Omega(a_1)\Omega(b_1)}\Big|
\end{equation}
where $a$ is the radial coordinate of the island as shown in Fig. \ref{ip} and it can be shown that $t_a=t_b=t$ extremizes the EE \cite{univer}:
\begin{align}
D^2(a_1,b_1)&=\Big|\bigg(\frac{e^{k(r^*_a+t_a)}}{k}-\frac{e^{k(r^*_b+t_b)}}{k}\bigg)\bigg(-\frac{e^{k(r^*_a-t_a)}}{k}+\frac{e^{k(r^*_b-t_b)}}{k}\bigg)\sqrt{\omega^2(a)\omega^2(b)f(a)f(b)e^{-2k(r^*_a+r^*_b)}}\Big|\\
&=\Big|\frac{1}{k^2}\Big[2e^{k(r^*_a+r^*_b)}\cosh\big(k(t_b-t_a)\big)-(e^{2kr^*_a}+e^{2kr^*_b})\Big]\sqrt{\omega^2(a)\omega^2(b)f(a)f(b)e^{-2k(r^*_a+r^*_b)}}\Big|
\end{align}

Then the EE is computed as follows:
\begin{align}
S_{rad.}&=\frac{A}{2G}+\frac{\mathcal{C}}{3}log\Big|\frac{1}{k^2}\Big[2e^{k(r^*_a+r^*_b)}-(e^{2kr^*_a}+e^{2kr^*_b})\Big]\sqrt{\omega^2(a)\omega^2(b)f(a)f(b)e^{-2k(r^*_a+r^*_b)}}\Big|\\
&=\frac{A}{2G}+\frac{2\mathcal{C}}{3}log\Big[\frac{e^{kr^*_a}-e^{kr^*_b}}{k}\Big]+\frac{\mathcal{C}}{6}log\bigg(\omega^2(a)\omega^2(b)f(a)f(b)e^{-2k(r^*_a+r^*_b)}\bigg)\label{iee}
\end{align}
To find the location of the island, we should extremize the above relation with respect to $a$ i.e. $\partial_{a}S_{rad.}=0$ which leads to:
\begin{align}
\frac{A'}{2G}&-\frac{2\mathcal{C}}{3}\frac{\frac{e^{kr^*_a}}{f(a)}}{\frac{e^{kr^*_a}-e^{kr^*_b}}{k}}
+\frac{\mathcal{C}}{6}\frac{1}{f(a)f(b)\omega^2(a)\omega^2(b)e^{-2k(r^*_a+r^*_b)}} \Big[f'(a)f(b)\omega^2(a)\omega^2(b)e^{-2k(r^*_a+r^*_b)}\nonumber\\
&-2kf(b)\omega^2(a)\omega^2(b)e^{-2k(r^*_a+r^*_b)}+f(a)f(b)2\omega(a)\omega'(a)\omega^2(b)e^{-2k(r^*_a+r^*_b)}\Big]=0
\end{align}
then we have
\begin{align}
\frac{3A'}{2G}=\mathcal{C}\Big[\frac{2ke^{kr^*_a}}{f(a)\big(e^{kr^*_a}-e^{kr^*_b}\big)}+\frac{2k-f'(a)}{2f(a)}-\frac{\omega'(a)}{\omega(a)}\Big]\;.\label{a}
\end{align}
If we suppose that $a$ is at the proximity of horizon $R$, $a\approx R$, then $f(a)\approx 2k (a-R)$ and $r^*(r)=\int^r\frac{dr}{f}\approx \int^r\frac{dr}{2k (r-R)}=\frac{1}{2k}log(r-R)$. The leading term in \eqref{a} is given by:
\begin{align}
&\frac{3A'}{2\mathcal{C}G}\approx \frac{2k (a-R)^{1/2}}{f(R) e^{kr^*_b}}=\frac{2k (a-R)^{1/2}}{f'(R)(a-R) e^{kr^*_b}}
\end{align}
It gives:
\begin{align} 
a=R+\Big(\frac{2e^{-kr^*_b}\mathcal{C}G}{3A'}\Big)^2\label{ar}
\end{align}
The solution \eqref{ar} shows that there is an island configuration for this model. By putting \eqref{ar} into the first term in \eqref{iee}, $\frac{A}{2G}$ and using Taylor expansion:
\begin{align}
\frac{A(a)}{2G}&\approx \frac{A(R)}{2G}+\frac{A'(R)}{2G} (a-R)\\
&=\frac{A(R)}{2G}+\frac{A'(R)}{2G} \Big(\frac{2e^{-kr^*_b}\mathcal{C}G}{3A'}\Big)^2\\
&\approx \frac{A(R)}{2G}+\frac{2e^{-2kr^*_b}\mathcal{C}^2G}{9A'(R)}\label{area}
\end{align}
To show that \eqref{iee} is equal to \eqref{anaee}, it is useful to show that by direct computation of \eqref{distance} there is a relation between $D^2(a,b)$ in the island prescription and $D^2(R,b)$:
\begin{align}
\frac{\mathcal{C}}{3}logD^2(a,b)-\frac{\mathcal{C}}{3}logD^2(R,b)&=\frac{2\mathcal{C}}{3}log\big(\frac{e^{kr^*_a}-e^{kr^*_b}}{e^{kr^*_b}-e^{kr^*(R)}}\big)+\frac{\mathcal{C}}{6}log\big(\frac{f(a)e^{-2kr^*_a}\omega^2(a)}{f(R)e^{-2kr^*(R)}\omega^2(R)}\big) \nonumber\\
&\approx \frac{\mathcal{C}}{6}log\big[\big(\frac{e^{kr^*_a}-e^{kr^*_b}}{e^{kr^*_b}}\big)^4\frac{f(a)}{f'(R)}e^{-2kr^*_a}\big]   \nonumber\\
&\approx \frac{\mathcal{C}}{6}log\big[1-4(a-R)^{1/2}e^{-kr^*_b}\big] \nonumber\\
&\approx \frac{\mathcal{C}}{6} \big(-4(a-R)^{1/2}e^{-kr^*_b}\big)=-\frac{4\mathcal{C}^2G e^{-2kr^*_b}}{9A'}
\end{align}
where we used approximations: 
\begin{align*}
f(a)&\approx f'(R)(a-R)\;, \quad f(R)e^{-2kr^*(R)}\approx f'(R)=2k\;,\quad
e^{2kr^*_a}\approx a-R\;,\quad  \omega^2(a)\approx \omega^2(R) 
\end{align*}

Now, putting \eqref{area}, \eqref{ar} into \eqref{iee}:
\begin{align}
S_{rad.}&=\frac{A(a)}{2G}+\frac{\mathcal{C}}{3}logD^2(a,b)  \nonumber\\
&=\frac{A(R)}{2G}+\frac{2e^{-2kr^*_b}\mathcal{C}^2G}{9A'(R)}+\frac{\mathcal{C}}{3}logD^2(R,b)-\frac{4\mathcal{C}^2G e^{-2kr^*_b}}{9A'} \nonumber\\
&=\frac{A(R)}{2G}-\frac{2\mathcal{C}^2G e^{-2kr^*_b}}{9A'}+\frac{\mathcal{C}}{3}log\Big(\frac{e^{2kr^*_b}}{k^2}\sqrt{f(b)f(R)e^{-2k(r^*_b+r^*(R))}\omega^2(b)\omega^2(R)}\Big)  \nonumber\\
&=\frac{A(R)}{2G}-\frac{2\mathcal{C}^2G e^{-2kr^*_b}}{9A'}+\frac{\mathcal{C}}{3}log\Big(\frac{e^{2kr^*_b}}{k^2}\sqrt{2kf(b)e^{-2kr^*_b} \omega^2(b)\omega^2(R)}\Big) \label{ltee}
\end{align}
For large $b$, we find (see appendix \ref{app:large-b} for details):
\begin{align}
S_{rad.}\approx \frac{\mathcal{C}}{6}log\Big(f(b) \omega^2(b)\Big) \label{appis}
\end{align}
which is the same as \eqref{app}. This implies that the island prescription is equivalent to inclusion of water makeup in the analogue gravity simulation of a black hole.

\section{Conclusion}
Information paradox received considerable attention because of the fact that defying quantum theory and the theory of general relativity. Recently, AdS/CFT through the duality provided a resolution to the paradox known as the island prescription. The idea originates from the finding the EE of the Hawking radiation via holographic proposals such as RT formula and its extensions. All the problem of finding the EE on the boundary is switched to the finding a surface in the bulk. The island prescription contains an additional surface after the Page time so-called the island and shows that its contribution to the EE of the Hawking radiation leads to the Page curve; in other words, the unitarity of the quantum theory. At first, this additional term seems to be ad hoc; however, the path integral computations in low dimensional gravities recovered the island term. Nonetheless, it is obscure how the quantum theory resolves the problem and it requires a quantum theory of gravity. However, looking at the problem in analogue gravity parlance and turning the paradox into a fluid problem can make a better understanding.

In this paper, we simulated the bulk theory of the Hawking radiation where the EE of the Hawking radiation is computed by finding the area of a surface with and without the island configuration. It showed that the information loss where there is no any island configuration is equivalent to the momentum loss of the fluid over the horizon of a dumb hole, and to maintain the horizon steadiness, we applied the Newton's second law of motion to the momentum and concluded that there should be a momentum inwards to the horizon. The water pump or any apparatus maintaining the momentum in real experiments is responsible for the maintenance of the inwards momentum flow, it could be a water pump, gravity in rivers that downs the water and etc. Taking into account the pump compensation led to the island prescription through the analogue gravity calculations, in other words after the Page time when there is a drop in the momentum flow rate, to make the horizon steady it requires an inwards momentum flow into the dumb hole. Then the emergence of the island is equivalent to the maintenance of the momentum. The interesting point is that if there would not the island prescription in holography, the analogue gravity calculations and the notion of the fluid could introduce the island prescription where the island configuration was not an ad hoc term, owing to the fact that the steadiness of the dumb hole horizon and Newton's second law necessitate the additional term. This opportunity opens up a new method or at least provides a glimpse of help to solve the problems of theoretical physics through the analogue gravity.

\section*{Acknowledgment}
Authors would like to appreciate Stefano Liberati and Xian-Hui Ge for their very useful comments. 
 This work was fully supported by the people  of Iran.

\appendix
\section{Analogue gravity: a review}\label{analogueapp}
To find solutions like black holes in fluids we can start by an irrotational fluid, Navier-Stokes and the continuity equations \cite{unruh}:
\begin{align}
&\nabla \times \vec{v}=0\\
&\rho \big(\frac{\partial \vec{v}}{\partial t}+(\vec{v}. \nabla)\vec{v}\big)=-\nabla p-\rho \nabla V\\
&\frac{\partial \rho}{\partial t}+\nabla . (\rho \vec{v})=0
\end{align}
where $V$ is potential energy, $p$ pressure and $\rho$ density. Change the variables:
\begin{align}
&\xi:=ln \rho\;,\qquad  \vec{v}:=\nabla \psi\\
&g(\xi):=\int^{e^\xi}\frac{1}{\rho'}\frac{dp(\rho')}{d\rho'}d\rho'
\end{align}
and introduce: 
\begin{equation}
\xi=\xi_0+\bar{\xi}\;,\qquad  \psi=\psi_0+\bar{\psi}
\end{equation}
then putting the new variables into Navier-Stokes equation to find linearized equation:
\begin{equation}
\frac{1}{\rho_0}\Big(\frac{\partial}{\partial t}\frac{\rho_0}{g'(\xi_0)}\frac{\partial \bar{\psi}}{\partial t}+\frac{\partial}{\partial t}\frac{\rho_0 \vec{v_0}}{g'(\xi_0)}. \nabla \bar{\psi}+\nabla .\big(\frac{\rho_0 \vec{v}}{g'(\xi_0)}\frac{\partial \bar{\psi}}{\partial t}\big)-\nabla .(\rho_0 \nabla \bar{\psi})+\nabla .\big(\frac{\vec{v} \rho_0 \vec{v}. \nabla \bar{\psi}}{g'(\xi_0)}\big)\Big)=0
\end{equation}
The above equation could be rewritten as follows:
\begin{equation}
\partial_{\mu}\big(\sqrt{-g} g^{\mu \nu}\partial_{\nu}\bar{\psi}\big)=0
\end{equation}
This is exactly the same as EoM of a massless scaler field in a background with metric:
\begin{equation}\label{met}
ds^2=\frac{\rho_0}{c(\rho_0)}\Big(\big(c^2(\rho_0)-v^2\big)dt^2+2dt \vec{v_0}. d\vec{v}-dx^2\Big)
\end{equation}
where:
\begin{equation}
c^2(\rho_0)=g'(ln \rho_0)
\end{equation}
\subsection{Conformal factor}
Eq. \eqref{met} shows that a massless scalar field in a curved background could be simulated by a fluid in a flat spacetime. However, simulating an arbitrary field theory in an arbitrary spacetime is not possible and we need a modification to the line element to make it feasible \cite{conformal}.
As a general method the analog metric could be written using the Lagrangian approach. Suppose the Lagrangian $\mathcal{L}$ of a fluid in $m+1$ dimensions is as follows:
\begin{align}
\mathcal{L}=\mathcal{L}(\eta^{\mu \nu} \partial_\mu \phi \partial_\nu \phi -V(\phi ,t,x))=\mathcal{L}(\mathcal{K} -V(\phi ,t,x))
\end{align}
The energy-momentum tensor by variational principle is given by:
\begin{align}
T_{\mu \nu}=-\Big(2\frac{\partial \mathcal{L}}{\partial \mathcal{K}}\partial_\mu \phi \partial _\nu \phi-\mathcal{L}\eta_{\mu \nu}\Big)
\end{align}
Note that the energy-momentum tensor for the fluid dynamics is read:
\begin{align}
T_{\mu \nu}=(p+\rho) u_\mu u_\nu +p \eta_{\mu \nu}
\end{align}
Comparing the two energy-momentum tensors, one can read the acoustic metric in terms of quantities in fluid dynamics:
\begin{align}
u_\mu&=\frac{\partial _\mu \phi}{\sqrt{\mathcal{K}}} \\
p&=\mathcal{L} \\
\rho&=2\mathcal{K}\frac{\partial\mathcal{L}}{\partial\mathcal{K}}-\mathcal{L}.
\end{align}
Perturbing the scalar field around a background $\phi=\phi_0+\epsilon \phi_1$, the Lagrangian up to the second order is as follows:
\begin{equation}
\mathcal{L}=\mathcal{L}_0+\epsilon \mathcal{L}_1+\epsilon^2\mathcal{L}_2+\mathcal{O}(\epsilon^3).
\end{equation}
By integrating by parts, the second order term would be:
\begin{align}
S=&S_0+\frac{\epsilon^2}{2}\int d^{m+1}x\Big(\big(\frac{\partial^2\mathcal{L}}{\partial(\partial_{\mu}\phi)\partial(\partial_{\nu}\phi)}\big)\partial_{\mu}\phi_1 \partial_{\nu}\phi_1+\big(\frac{\partial^2\mathcal{L}}{\partial\phi \partial\phi}-\partial_{\mu}(\frac{\partial^2\mathcal{L}}{\partial(\partial_{\mu}\phi) \partial\phi})\big)\phi_1\phi_2\Big)+ \mathcal{O}(\epsilon^3)  \nonumber\\
:=&S_0+\frac{\epsilon^2}{2}\int d^{m+1}x \Big(\sqrt{-g}g^{\mu\nu}\partial_{\mu}\phi_1 \partial_{\nu}\phi_1-\sqrt{-g} m_{eff}^2 \phi_1^2\Big)
\end{align}
Applying chain rule leads to:
$$\frac{\partial}{\partial(\partial_{\mu}\phi)}=\frac{\partial \mathcal{K}}{\partial(\partial_{\mu}\phi)}\frac{\partial}{\partial\mathcal{K}}=2\eta^{\mu\nu}\partial_{\nu}\phi \frac{\partial}{\partial\mathcal{K}}$$
the term $\sqrt{-g}g^{\mu\nu}$ is written as:
\begin{align}
&\sqrt{-g}g^{\mu\nu}=-2\Big(\eta^{\mu\nu}\frac{\partial\mathcal{L}}{\partial \mathcal{K}}-2\mathcal{K}u^{\mu}u^{\nu}\frac{\partial^2\mathcal{L}}{\partial \mathcal{K}^2}\Big). \label{dm}\\
&\sqrt{-g}m_{eff.}^2=-\frac{\partial^2\mathcal{L}}{\partial \phi \partial \phi}+\partial_{\nu}\Big(\frac{\partial^2\mathcal{L}}{\partial(\partial_{\nu}\phi)\partial \phi}\Big)
\end{align}
Defining the speed of sound $c^{-2}=\frac{\partial \rho}{\partial p}$.
Equation of motion is read as:
\begin{equation}
\Box \phi_1-m_{eff}^2\phi_1=\frac{1}{\sqrt{g}}\partial_{\mu}\Big(\sqrt{|g|}g^{\mu\nu}\partial_{\nu}\phi_1\Big)-m_{eff}^2\phi_1=0\label{eom}
\end{equation}
In the rest frame $u_{\mu}=(1,\vec{0})$ and by use of eq. \eqref{dm}, the determinant of $g_{\mu \nu}$ is given by:
\begin{equation}
\sqrt{-g}=c^{-\frac{2}{m-1}}\Big(-\frac{\rho+p}{\mathcal{K}}\Big)^{\frac{m+1}{m-1}}
\end{equation}
The acoustic metric is read:
\begin{align}
g_{\mu \nu}=c^{\frac{-2}{m-1}}\Big(\frac{\rho+p}{\mathcal{K}}\Big)^{\frac{2}{m-1}}\Big(\eta_{\mu \nu}+(1-c^2)u_\mu u_\nu\Big) \label{am}
\end{align}
By rescaling the metric $\bar{g}=\omega^{-2}g$ in \eqref{eom}, the d'Alembertion would be:
\begin{equation}
\bar{\Box} \phi_1=\frac{1}{\omega^n\sqrt{|\bar{g}|}}\partial_{\mu}\Big(\omega^{n-2}\sqrt{|\bar{g}|}\bar{g}^{\mu\nu}\partial_{\nu}\phi_1\Big)
\end{equation}
and rescaling $\bar{\phi}_1=\omega^{\frac{n-2}{2}}\phi_1$, \eqref{eom} is rewritten as:
\begin{equation}
\bar{ \Box}\bar{\phi}_1-\bar{m}_{eff}^2\bar{\phi}_1=0\label{bareom}
\end{equation}
where:
\begin{equation}
\bar{m}_{eff}^2=\omega^2 m_{eff}^2+\omega^{\frac{2-n}{n}}\bar{\Box}\omega^{\frac{n-2}{2}}
\end{equation}
Eq. \eqref{bareom} shows that Weyl transformation of an analogue model could be constructed at the expense of a new effective mass of the field.

\section{Large $b$ approximation}\label{app:large-b}

In deriving \eqref{appis}, we use large $b$ approximation, and conduct the computations for an AdS-Schwarzschild black hole in $d+1$ dimensions:
\begin{equation}
f(r)=l+\frac{r^2}{L^2}-\frac{\Omega^{d-2}}{r^{d-2}}
\end{equation}
where $\Omega^{d-2}=R^{d-2}\big(\frac{R^2}{L^2}+l\big)$ and $L$ the AdS radius. By coordinate transformation $z=\frac{L^2}{r}$, $\omega$ for the black hole with flat horizon, $l=0$\footnote{In 3+1 dimensions $r^*$ is analytically computed when $l=0$, so we go on with this choice.} in 3+1 dimensions where we compare them to its analogue for large $b$ is given:
\begin{align}
&ds^2=-\frac{\bar{\omega}^2L^2}{z^2}\bigg(\Big(1-\big(\frac{z}{z_0}\big)^{d}\Big)dt^2+\frac{dz^2}{\Big(1-\big(\frac{z}{z_0}\big)^{d}\Big)}+\sum_{i=1}^{d-1}dx_i^2\bigg) \;, \label{flat}\\
&f(z)=1-\big(\frac{z}{z_0}\big)^{d}\;.
\end{align}
By coordinate transformation $t=t'\alpha+\beta$ for $\beta=\int dz\sqrt{1-f}/f $, \eqref{flat} could be written as:
\begin{align}
ds^2=\omega^2\Big(-\alpha^2fdt'^2+2\alpha\sqrt{1-f}dt'dz+dz^2+\sum_{i=1}^{d-1}dx_i^2\Big)
\end{align}
where $\omega:=\frac{\bar{\omega} L}{z}$, then the fluid parameters are written as \eqref{vel} and \eqref{den}. To satisfy the continuity equation:
\begin{align}
&\nabla.(\rho v)=0 \rightarrow \partial_z (\rho v)=0\\
&\omega=\frac{\gamma}{\kappa^{\frac{1}{n-1}}}(1-f)^{\frac{1}{2(1-n)}}=\frac{\gamma}{\kappa^{\frac{1}{2}}}(\frac{z}{z_0})^{-\frac{3}{4}}
\end{align}
where we set $d=n=3$. The large $b$ limit means $z\ll 1$ and $\omega (b) \gg 1$, then in \eqref{ltee}:
\begin{align}
z^*(b)&=\int\frac{dz}{1-\big(\frac{z}{z_0}\big)^3}=\frac{z_0}{6} \left(\log \left(z_0^2+z_0 z+z^2\right)-2 \log (z_0-z)+2 \sqrt{3} \tan ^{-1}\left(\frac{z_0+2 z}{\sqrt{3} z_0}\right)\right) \nonumber\\
&\approx \frac{\sqrt{3}\pi z_0}{18}\\
&log\Big(\sqrt{f(b)\omega^2(b)}\Big) \gg log\Big(\frac{e^{2kr^*_b}}{k^2}\sqrt{e^{-2kr^*_b}2k \omega^2(R)}\Big)
\end{align}
so \eqref{appis} is recovered.


\end{document}